\documentclass[prd,showpacs,nofootinbib]{revtex4}
\usepackage{graphicx, epsfig}
\usepackage{color}
\usepackage{mathrsfs}
\usepackage {amssymb}
\usepackage {amsmath}

\input epsf
\usepackage{color,graphics,latexsym,amsfonts}

\begin{document}
\title{Relativistic stars in  f(R)  and scalar-tensor theories}
\author{Eugeny Babichev$^{a,b}$, David Langlois$^{a,c}$}
\affiliation{$^a$ APC, UMR 7164 (CNRS-Universit\'e Paris 7), 10 rue Alice Domon et L\'eonie Duquet,75205 Paris Cedex 13, France}
\affiliation{$^b$ Institute for Nuclear Research of the Russian Academy of Sciences, 
60th October Anniversary Prospect, 7a, 117312 Moscow, Russia}
\affiliation{$^c$ IAP (Institut d'Astrophysique de Paris), 98bis Boulevard Arago, 75014 Paris, France}
\begin{abstract}
We study relativistic stars in the context of  scalar tensor theories of gravity that try to account for the observed cosmic acceleration and satisfy the 
local gravity constraints via the chameleon mechanism.  More specifically, we consider two types of models: scalar tensor theories with an inverse 
power law potential and f(R) theories. Using a relaxation algorithm, we construct numerically static relativistic stars, both for  
constant energy density configurations and for a polytropic equation of state.  
We can reach a gravitational potential up to $\Phi\sim 0.3$ at the surface of the star, even in f(R) theories with an ``unprotected" curvature singularity.
However, we find static configurations only if the pressure does not exceed one third of the energy density, except possibly in a limited region of 
the star (otherwise, one expects tachyonic instabilities to develop). 
This constraint is  satisfied by realistic equations of state for neutron stars.
\end{abstract}


\date{\today} 
\maketitle

\def\beq{\begin{equation}}
\def\eeq{\end{equation}}
\def\be{\begin{equation}}
\def\ee{\end{equation}}
\newcommand{\bea}{\begin{eqnarray}}
\newcommand{\eea}{\end{eqnarray}}

\def\mP{{M_P}}  % reduced Planck mass=1/\sqrt{8\pi G}
\def\c{{\rm c}}

\newcommand{\bse}{\begin{subequations}}
\newcommand{\ese}{\end{subequations}}

\def\a{{\alpha}}
\def\d{{\delta}}
\def\e{{\epsilon}}
\def\tg{{\tilde g}}
\def\tR{{\tilde R}}
\def\tH{{\tilde H}}
\def\tT{{\tilde T}}
\def\tn{{\tilde n}}
\def\trho{{\tilde\rho}}
\def\tP{{\tilde P}}
\def\hphi{{\hat\phi}}
\def\hrho{{\hat \rho}}
\def\hP{{\hat P}}
\def\hV{{\hat V}}
\def\hm{{\hat m}}
\def\ms{M_*}
\def\rs{r_*}
\def\hM{{\hat{\cal M}}}
\def\F{{\cal F}}
\def\V{{\cal V}}
\def\dphi{{\delta\phi}}
\def\phim{{\phi_{\rm min}}}

\def\R0{{R_0}}
\def\x1{x_\infty}
\def\Om{\Omega}
\def\as{\infty}

\par\bigskip
\section{Introduction}
One of the most challenging tasks for cosmology and fundamental physics today is to try to understand the apparent acceleration of the Universe. 
Beyond the minimal assumption of a pure cosmological constant, two main approaches have been explored. The first one consists in assuming 
some unknown form of matter, called {\it dark energy}, characterized by an equation of state  $P\simeq -\rho$. The second approach is more 
radical, as it tries to explain the present observations as the manifestation of a modified theory of gravity, which mimicks general relativity on solar 
system scales, but significantly deviates from it on cosmological scales. 
 
 It turns out that it is rather difficult to construct a theory of gravity that, while being  internally consistent, can account for the cosmological 
observations and be compatible with the present gravity constraints deduced from laboratory experiments and from  solar  and astrophysical 
systems.
 A class which has attracted a lot of attention is the so-called $f(R)$ gravity theories where the standard Einstein-Hilbert gravitational Lagrangian, 
proportional to the scalar curvature $R$, is replaced by a function of $R$ while the matter part of the Lagrangian is left unchanged (see e.g. 
\cite{review} for a recent review). 
After several detours, it has been realized that viable $f(R)$ theories must satisfy stringent conditions in order to avoid instabilities and to satisfy the 
present laboratory and solar system constraints, and a few models have been carefully constructed to meet these requirements 
\cite{Hu:2007nk,Appleby:2007vb,Starobinsky:2007hu}.

To explore the full viability of these theories, it is important to go one step further by studying their behaviour  in the  strong gravity regime, such as 
reigns in the core of the most relativistic stars, namely neutron stars. In this context, it has been claimed in \cite{Frolov,KM1} that  very relativistic 
stars do not exist  in the models  \cite{Hu:2007nk,Appleby:2007vb,Starobinsky:2007hu} because of the presence of an easily accessible 
singularity. 
In a recent work \cite{BL1}, we have shown that this claim does not hold by constructing numerically static relativistic stars. 
Note that  relativistic stars have also been studied in \cite{Kainulainen:2007bt}  but within  different $f(R)$ models.

$f(R)$ models can also be seen as a subclass of scalar-tensor theories. In particular, viable $f(R)$ models rely on the so-called chameleon 
mechanism. For this reason,  it is interesting to extend the study of relativistic stars to  chameleon models.
 In this work, we show that the behaviour of chameleon and $f(R)$ models is quite similar. In particular, in both cases, the scalar field in the 
innermost part of the star sticks to the minimum of its effective potential  (if  this minimum exists). If the equation of state is such that  $\rho-3P<0$, 
which occurs in the central part of  highly relativistic constant energy density stars, then there is no minimum. As we show here, it is nevertheless 
possible to construct numerically static stars, up to some critical value of the central energy density. 
We believe that this is due to tachyonic instabilities, associated with a negative effective squared mass, which develop and prevent the existence 
of a static star configuration. 
However, this problem does not apply to  realistic neutron stars: although there is a large uncertainty on the equation of state deep inside a 
neutron star, the equations of state that have been proposed in the literature verify $\rho-3P>0$ throughout the star. To construct   a simple 
approximation of a realistic neutron star, we  have used a polytropic equation of state.

{ Although, according to our previous work \cite{BL1} and the present one, the singularity of the models 
\cite{Hu:2007nk,Appleby:2007vb,Starobinsky:2007hu} does not seem so far to be an obstacle  for relativistic stars, it appears to be problematic for cosmology~\cite{Frolov}.}  
This has motivated 
the construction of regularized versions of these models by adding for instance an extra $R^2$ term~
\cite{Starobinsky:2007hu,Capozziello:2009hc,Dev:2008rx,Abdalla:2004sw,Thongkool:2009js}. A more sophisticated model was also proposed recently in~\cite{Appleby:2009uf}.  We 
will consider these ``cured''  $f(R)$ theories and compare their behaviour in relativistic stars with their ``singular'' counterparts.

This paper is organized as follows. In the next section, we derive the main equations governing static and spherically  symmetric configurations in 
scalar-tensor theories. Section III is devoted to relativistic stars in chameleon models. We then consider, in Section IV, the $f(R)$ models, including 
the ``cured'' models that have recently been introduced to solve the cosmological singularity problem of some of these models. We finally 
conclude in Section V.

\section{Static and spherically symmetric equations}
In the present work, we consider models characterized by an action of the form
\beq
\label{action}
S= \int d^4x \sqrt{-g}\left[\frac{\mP^2}{2}{R}-\frac{1}{2}\left({\nabla}\phi\right)^2-V(\phi)\right]  +{S}_m (\Psi_m; \tg_{\mu\nu})\,,
\eeq
with $M_P^2\equiv 1/(8\pi G)$, and
where the matter (i.e. the fields $\Psi_m$) is minimally coupled to the metric 
\beq
\tg_{\mu\nu}=\Om^2(\phi) g_{\mu \nu}.\nonumber
\eeq
For simplicity, we  consider only couplings of the form
\beq
\label{a}
\Om=\exp{\left(Q\frac{\phi}{M_P}\right)},
\eeq
where $Q$ is a constant, although more general functions can have interesting effects (such as the spontaneous scalarization dicovered in 
\cite{Damour:1993hw}
for coupling functions of the form $\ln\Om=\beta\phi^2/2$ with $\beta<0$).

Viable scalar-tensor theories are severely restricted by the present constraints on gravity, coming from laboratory experiments, solar system and 
binary pulsar tests. For example, deviations from General Relativity (GR) can be parametrized by the post-Netwonian parameters $(\tilde\beta-1)$ 
and $(\tilde\gamma-1)$. For the coupling (\ref{a}), one finds (see e.g. \cite{Damour:1992we}\footnote{Note that our $\phi$ is related to the scalar 
field $\varphi$ defined in \cite{Damour:1992we}
by $\phi=\sqrt{2} M_P\varphi$ and therefore their $\alpha$ corresponds to $\alpha=\sqrt{2}Q$.}) 
\beq
\tilde\beta-1=0, \qquad 
\tilde\gamma-1=-4 \frac{Q^2}{1+2Q^2}\, .\nonumber
\eeq
The present   constraint, inferred from  solar system tests, is \cite{Bertotti:2003rm} 
\beq 
\mid \tilde\gamma-1\mid\lesssim 2\times 10^{-5}.\nonumber
\eeq
The binary pulsars also give a similar constraint, but which is  so far   weaker than the solar system constraint in our case\footnote{This can be 
seen on the figure 11 of the very recent lecture notes by G. Esposito-Farese \cite{EspositoFarese:2009ta}. For models with a  coupling of the form 
(\ref{a}), the parameter space is restricted to the vertical axis $\beta_0=0$ where, so far, the solar system constraints turn out to be more stringent.}.
The above constraint suggests that scalar tensor theories of gravity can be viable only for very small couplings between ordinary matter and the 
scalar field. 

However, this constraint can be evaded if the scalar field is endowed with a potential such  that its effective mass becomes large in the presence of 
matter. This is the so-called  chameleon effect ~\cite{Khoury:2003aq,Khoury:2003rn}. As a consequence of this effect, the ``bare" coupling constant 
$Q$ is replaced by an effective coupling $Q_{\rm eff}$ which can be  strongly suppressed, i.e. such that $|Q_{\rm eff}|\ll |Q|$. The purpose of this 
paper is to study relativistic stars in this type of  gravity. 
This also applies to   $f(R)$ theories, which  can be recast in the form of scalar tensor theories with   $Q=-1/\sqrt{6}$,  leading to the 
unacceptable value $\tilde\gamma=1/2$ \cite{Chiba:2003ir}. Therefore, these theories can be considered 
as viable  if the corresponding potential allows for a chameleon effect  \cite{Navarro:2006mw,Faulkner:2006ub,Brax:2008hh}.

\subsection{Links between the Jordan and Einstein frames}
The action (\ref{action}) is defined in the Einstein frame, where gravity is described by the usual Einstein-Hilbert term for the  metric $g_{\mu\nu}$. 
This metric differs from the metric  $\tilde{g}_{\mu\nu}$, defined in the Jordan frame, which is directly felt by the matter. As a consequence, one 
must be careful to specify the metric with respect to which various quantities are defined. For example, 
the energy-momentum tensor, defined in the  Jordan frame and denoted $\tT_{\mu\nu}$, is related to the energy-momentum tensor defined in the 
Einstein frame,
$T_{\mu\nu}$,  by
\beq
\tT^\mu_\nu=\Om^{-4} T^\mu_\nu,  \nonumber
\eeq
so that, for a perfect fluid, we have
\beq
\rho=\Om^4\trho, \quad P=\Om^4\tP.\nonumber
\eeq
The equation of state will always be specified in the Jordan frame.
When  $|Q\phi/M_P|\ll 1$, as will be the case for our relativistic stars, the numerical values for the energy density and the pressure are essentially 
the same in the two frames.

\subsection{Equations of motion}
In Einstein's frame, the equations of motion are given by 
\bea
\label{einstein}
G_{\mu\nu}&=& M_P^{-2} \left[ T_{\mu\nu}^{(m)}+\partial_\mu\phi\partial_\nu\phi-\frac12 g_{\mu\nu}\partial^\sigma\phi\partial_\sigma\phi-Vg_{\mu
\nu}\right]\, ,\\
\label{Klein-Gordon}
\nabla_\sigma\nabla^\sigma \phi&=&-\frac{dV}{d\phi}-\frac{\Omega'}{\Omega}T^{(m)}, 
\eea
where $G_{\mu\nu}\equiv R_{\mu\nu}-(1/2) R \, g_{\mu\nu}$ is the Einstein tensor,  $T_{\mu\nu}^{(m)}$  is the energy-momentum tensor for the fluid 
matter of the star (i.e. does not include the scalar field) and
\beq
T^{(m)}\equiv g^{\mu\nu}T^{(m)}_{\mu\nu}=-\rho+3P 	\nonumber
\eeq
is its trace.

We now consider a static and spherically symmetric geometry, with metric 
\beq
ds^2=-e^\nu dt^2+e^\lambda dr^2+r^2 \left(d\theta^2+\sin^2\theta\, d\phi^2\right).	\nonumber
\eeq
Introducing the radial function $m(r)$
so that 
\beq
 e^{-\lambda}\equiv1-2m /r, 	\nonumber
\eeq
the time and radial components of Einstein's equations (\ref{einstein}) yield, respectively,  

\bse
\begin{align}
\label{tt}
m' & =\frac{r^2}{2M_P^2}\left[\Omega^4 \trho+\frac12 e^{-\lambda}{\phi'}^2+V(\phi)\right],\\
\label{rr}
\nu'& =e^\lambda\left[\frac{2m}{r^2}+\frac{r}{M_P^2}\left(\frac12 e^{-\lambda}{\phi'}^2-V(\phi)\right)+
\frac{r\Omega^4 \tP}{M_P^2}\right],
\end{align}
\ese
where one recognizes, inside the brackets of the first equation,  the total energy density, which includes the fluid energy density (defined in the 
Einstein frame) as well as the gradient and potential energies of the scalar field. 

Instead of the angular component of Einstein's equations, it is convenient to use  the conservation of the fluid energy-momentum, which has the 
standard form,
\beq
\tilde\nabla_\mu \tT^{\ \mu}_{(m) \nu}=0,\nonumber
\eeq
in the Jordan frame, where the matter is minimally coupled. 
This gives 
\beq
\label{Bianchi}
\tP'=-\frac12\left(\trho+\tP\right)\left(\nu'+2\frac{\Omega'}{\Omega}\phi'\right).
\eeq
The last equation is provided by   the Klein-Gordon equation, Eq.~(\ref{Klein-Gordon}), for the scalar field,
\beq
\label{KG}
\phi''+\left(\frac{2}{r}+\frac12(\nu'-\lambda')\right)\phi'=e^\lambda\left[\frac{dV}{d\phi}+\Omega^3 \Omega'(\trho-3\tP)\right].
\eeq
Finally,  an equation of state,
\be
\label{EoS}
\tP=\tP(\trho),
\ee
closes the system of equations (\ref{tt}), (\ref{rr}), (\ref{Bianchi}) and (\ref{KG}).

\subsection{Constant energy density stars}
In general relativity, one can solve analytically the profile of a relativistic star by assuming that the energy density is constant,
\beq
  \rho=\rho_0\, .\nonumber
  \eeq
Using (\ref{tt})   in the GR limit, which  corresponds to $\phi=0$ and $V(\phi)=0$,
this implies   
  \beq
  m(r)=\frac{\rho_0 r^3}{6 M_P^2}\, .\nonumber
  \eeq
Substituting in the Tolman-Oppenheimer-Volkov equation
\beq
P'=-\frac{\rho+P}{r^2(1-2m/r)}\left[m+\frac{r^3 P}{2M_P^2}\right], \nonumber
\eeq
which follows from (\ref{Bianchi}) and (\ref{rr}) in the GR limit, one finds, 
 after integration, that the pressure profile is  given by  
  \beq
  \label{P_constant}
    P(r)=\rho_0 \frac{\left(1-{\frac{2G\ms}{r_*}}\right)^{1/2}-\left(1-{\frac{2G\ms r^2}{r_*^3}}\right)^{1/2}}
    {\left(1-{\frac{2G\ms r^2}{r_*^3}}\right)^{1/2}-3 \left(1-\frac{2G\ms}{r_*}\right)^{1/2} }\,
  \eeq
where $\rs$ is the radius of the star and $\ms\equiv 4\pi \rho_0 \rs^3/3$ its mass.

We can use the above analytical results to compute the trace of the energy momentum-tensor, which  is the quantity that couples directly to the 
scalar field in its equation of motion. The relation 
 (\ref{P_constant}) implies
 \beq
 \rho-3 P=  \rho_0 \frac{4\left(1-{\frac{2G\ms r^2}{r_*^3}}\right)^{1/2}-6\left(1-{\frac{2G\ms}{r_*}}\right)^{1/2}}
  {\left(1-{\frac{2G\ms r^2}{r_*^3}}\right)^{1/2}-3 \left(1-{\frac{2G\ms}{r_*}}\right)^{1/2} }. \nonumber
\eeq
Since this is a function that grows with increasing $r$, its minimal value is at the center of the star and is given by
\beq
\rho_c-3 P_c=\rho_0 \frac{4-6\left(1-2\Phi_*\right)^{1/2}}
 {1-3 \left(1-2\Phi_*\right)^{1/2} },\qquad \Phi_*\equiv \frac{G\ms}{\rs},\nonumber 
\eeq
where $\Phi_*$ is the gravitational potential at the star surface.
Therefore, whenever the compactness of the star is higher than the critical value
\beq
\label{Phi_critical}
\Phi_*=\frac{5}{18}\simeq 0.28,
\eeq
the quantity $\rho-3P$ becomes negative in the most central layers of the star. As we will see later, the  sign of $\rho-3P$ is crucial to understand  
the behaviour of the scalar field inside the star and can have drastic implications concerning the stability of the scalar field. Note that this critical 
value was also pointed out in \cite{Harada:1997mr}. Note also that although our derivation above applies only to constant energy density stars in 
pure general relativity, it remains a good approximation  for the stars we consider because the backreaction of the scalar field on the star profile is 
very small.

\subsection{More realistic equation of state}
Although analytically simple, a constant energy density star is not a very realistic substitute for a real neutron star. 
Since the equation of state  deep inside  a neutron star is  still unknown, we use
 in this paper a polytropic equation of state, which remains simple and is believed to represent a reasonable approximation for a real neutron star.  
Following \cite{EoSNS} we use,
\beq
\label{eos}
\trho(\tn)=m_B\left( \tn+K\frac{\tn^2}{n_0}\right) , \quad \tP(\tn)=K m_B\frac{\tn^2}{n_0},
\eeq
with $m_B=1.66\times 10^{-27}$ kg, $n_0=0.1\, {\rm fm}^{-3}$ and $K=0.1$.  
\begin{figure}[ht]
\centering
\includegraphics[width=0.5\textwidth, clip=true]{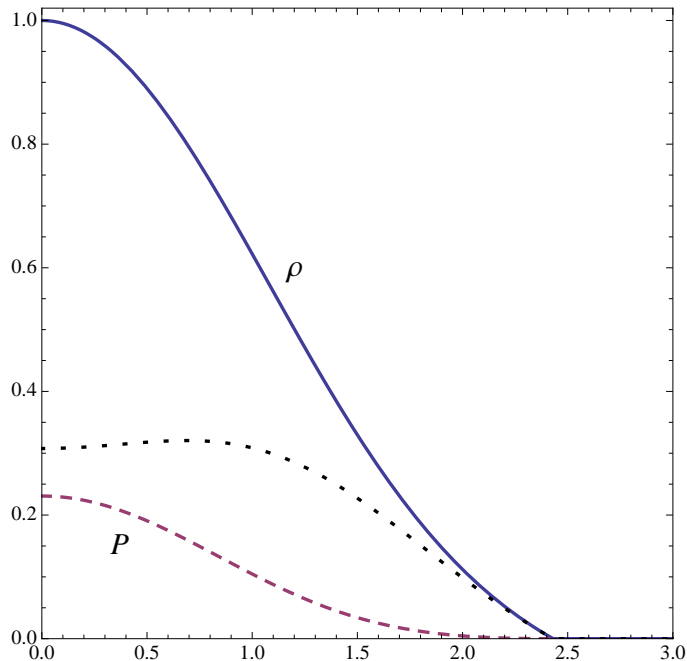}
\caption{Energy density $\trho$ (solid blue line), pressure $\tP$ (dashed purple line) and the combination $\trho-3\tP$ (black dotted line), 
in units of the central density $\rho_c$,
as functions of the radial coordinate $r$ (in units of $M_P\tilde{\rho}_c^{-1/2}$).}
\label{star}
\end{figure}
We take for the central particle number density the value $\tn_c = 0.3 \, {\rm fm}^{-3}$, which corresponds to a central energy density $\trho_c=6.47\times 10^{14}$~g/cm$^3$. In general relativity, the numerical integration of the Tolman-Oppenheimer-Volkov equations leads to a star, which is stable, with a global mass $M=3.10 M_\odot$ and a gravitational potential $|\Phi_*|\simeq 0.21$.
In both the chameleon and $f(R)$ models, we will use the same equation of state with the same central density and will find that the star configuration is almost identical to the general relativistic one.

In Fig.~\ref{star}, we plot the radial profile of the energy density and of the pressure. 
As can be seen in the same figure, the quantity   $\trho-3\tP$ remains positive throughout the star. The same  holds for other  realistic equations of 
state that have been proposed in the literature (see e.g. \cite{Lattimer:2006xb} and references therein for a discussion on the equations of state of 
neutron stars).
Interestingly, $\trho-3\tP$ is not a monotonous function of  the radius in the example we have chosen. We will see that this has a visible influence 
on the profile of the scalar field inside the star.

\subsection{Effective potential}

Assuming some  configuration for the relativistic star and ignoring any backreaction of the 
scalar field on this configuration, it is useful to interpret the right hand side of the Klein-Gordon equation (\ref{KG}), up to the metric component $e^
\lambda$, as the derivative of an effective potential defined by
\beq
\label{Veff}
V_{\rm eff}=V+\frac14 \Om^4 (\trho-3\tP).
\eeq
Consequently, a local extremum of the potential is characterized by the condition 
\beq
\label{dVeff}
\frac{d V_{\rm eff}}{d\phi}=\frac{dV}{d\phi}+\frac{Q}{\mP}e^{4Q\phi/\mP}(\trho-3\tP)=0.
\eeq
A solution of  the above equation, if it exists,  corresponds to a minimum of the potential if the effective squared mass
\beq
\label{meff2}
m_{\rm eff}^2\equiv \frac{d^2 V_{\rm eff}}{d\phi^2}= \frac{d^2V}{d\phi^2}+4\frac{Q^2}{\mP^2}e^{4Q\phi/\mP}(\trho-3\tP)\nonumber
\eeq
is positive. It is thus manifest that the quantity $\trho-3\tP$ plays a crucial r\^ole in the overall sign of $m_{\rm eff}^2$, with consequences for the 
stability of the star, which we discuss just below.

%%%%%%%%%%%%%%%%%%%%%%%%%%%%%%%%%%%%%
\subsection{Stability}
Although the study of the stability of our star configurations would
require a detailed analysis, which we leave for  future work, one can try
to discuss this issue with some qualitative arguments. For simplicity, we
ignore the perturbations of the star matter and of the scalar field and
consider only  the perturbations of the scalar field about the background
configuration. Decomposing these  perturbations into spherical harmonics,
\beq
\delta\phi(t,r,\theta,\phi)=\sum \delta\phi_{lm}(t,r) Y_{lm}(\theta,\phi), \nonumber
\eeq
one finds that each mode verifies the equation
\beq
\ddot\dphi-e^{\nu-\lambda}\left[\dphi''+\left(\frac{\nu'-\lambda'}{2}+\frac{2}{r}\right)\dphi'\right]+
e^{\nu}\left[\frac{l(l+1)}{r^2}+m^2_{\rm eff}\right]\dphi=0.	\nonumber
\eeq
where $m_{\rm eff}^2$ has been defined in (\ref{meff2}).
Using a Fourier decomposition in time, and denoting  the frequency $\omega$, the above equation of motion leads to a dispersion relation of the form
\beq
\omega^2\simeq k^2  +m^2_{\rm eff}, 	\nonumber
\eeq
where $k^{-1}$ is the typical lengthscale of variation of the scalar field. One can write  $k=\alpha r_*^{-1}$, where $\alpha$ is typically bigger than 
$1$ but not very much, in the situations we consider. Let us also assume that the dominant contribution to $m_{\rm eff}^2$ comes from the matter 
part.  Concentrating on the radial modes (which are potentially more dangerous), we find that 
the configuration is stable if,
\beq
\omega^2\simeq \frac{\alpha^2}{r_*^2}+4\frac{Q^2}{M_P^2}\trho_*(1-3w)>0,\nonumber
\eeq
{ where}
\be
w\equiv \tP/\trho.    \nonumber
\ee
Substituting $\trho_*\sim M_*/ r_*^3$ and $\Phi_*\sim M_*/(\mP^2 r_*)$ leads to the condition
\beq
(1-3w)\Phi_*\gtrsim-1, \nonumber
\eeq
where the right hand side is some constant, but typically of order $-1$. According to this relation, obtained in a very crude way, one expects 
instabilities to appear whenever the equation of state is such that $1-3w<0$ {\it and} the star is very compact so that $\Phi_*$ is a significant 
fraction of one. This is precisely the situation which one encounters for very massive stars with constant energy density, where $w>1/3$ in the 
central region.

\subsection{Numerical procedure}
{We now present how we proceed numerically. As a fist step, it is convenient to rescale the various quantities involved and to work directly with 
dimensionless quantities. Let us thus  introduce the following rescaled variables
\beq
\label{rescaling}
r=r_0\xi,\quad \phi=M_0\hphi,\quad V=\frac{M_0^2}{r_0^2}\hV,\quad \trho=\frac{M_1^2}{r_0^2}\hrho,\quad \tP_*=\frac{M_1^2}{r_0^2}\hP,
\eeq
where $r_0^{-1}$, $M_0$, $M_1$ are  parameters of dimension of mass, which are so far arbitrary. 
Substituting the above rescalings (\ref{rescaling}) in the equations of motion (\ref{tt}), (\ref{rr}), (\ref{Bianchi}) and (\ref{KG}), 
we obtain the following system of equations,
\bse
\label{eqs-r}
\begin{align}
\frac{2\hM'}{\xi^2}&= \e_1^2 \hrho e^{4Q\e_0\hphi}  + \e_0^2\left(\frac12 e^{-\lambda}{\hphi}'^2+\hV\right), \\
\nu' e^{-\lambda} &=  \frac{2\hM}{\xi^2}  + \e_1^2 \xi \hP e^{4Q\e_0\hphi} +\e_0^2\xi\left(\frac12 e^{-\lambda}{\hphi}'^2-\hV \right),\\
\hP' &=  -\frac12\left(\hrho+\hP\right)\left(\nu'+2 Q \e_0 \hphi'\right),\\
0 &=  \hphi''+\left(\frac{2}{\xi}+\frac12(\nu'-\lambda')\right)\hphi' 
	- e^\lambda\left[\frac{d\hV}{d\hphi}+\frac{\e_1^2}{\e_0}Q e^{4Q\e_0\hphi}(\hrho-3\hP)\right] ,
\end{align}
\ese
where we have used the particular  coupling (\ref{a}) and introduced the notations,
\bea
\hM\equiv\frac12 \xi\left(1-e^{-\lambda}\right),\qquad \e_0\equiv\frac{M_0}{M_P},\qquad \e_1\equiv\frac{M_1}{M_P}\, .\nonumber
\eea
The values of the parameters $\epsilon_0$ and $\epsilon_1$ have no physical relevance. They are chosen purely for convenience. 
In the following, we will  take  $\epsilon_1=1$ in all cases, but our choice for $\epsilon_0$ will be different for chameleon models and for $f(R)$ gravity.

Meanwhile the equation of state (\ref{EoS}) transforms into 
\be
\label{eos-r}
\hP=\hP(\hrho).
\ee
For the polytropic equation of state Eq.(\ref{eos}), with $\tilde n_c=0.3$ fm$^{-3}$,  we will always choose
$r_0=M_P \trho_\c^{-1/2}$,
so that 
\beq
\hrho_\c=1 \qquad ({\rm polytropic \ star}). \nonumber
\eeq
The  equation of state Eq.(\ref{eos}) then becomes, in these rescaled units,
\beq
\hrho=\hP+\sqrt{\frac{\hP}{0.39}}\quad \iff \quad \hP = \frac{1}{1.56}\left(\sqrt{1+1.56\hrho}-1\right)^2,\label{eos-R} 
\eeq
with  the rescaled central pressure  $\hP_c \simeq 0.23$.
Since the neutron star is 
a highly relativistic object, this choice  entails that the radius of the star in 
rescaled units is of order one, $\xi_*\equiv r_*/r_0\sim 1$.

Since we study static configurations,  a natural way to proceed with numerics is to use a relaxation algorithm. In the following calculations the 
relaxation parameter will be taken to vary with the iteration number (typically between $10^{-3}$ and $1$), in order to ensure the convergence of 
the algorithm and to minimize  the number of iterations. We have used a non-homogenous grid, usually between 1000 and  6000 points. 
The details of the grid depend on the  model under consideration and its parameters. 
Normally the grid is chosen so that the points of the grid are highly concentrated in the particular region inside the star where the solution varies 
rapidly. Such a choice of inhomogeneous grid allows us to numerically resolve potentially problematic regions.

Let us finally discuss the boundary conditions. The system (\ref{eqs-r}), which includes three first-order differential equations and one second-order 
differential equation, requires five boundary conditions. Some of them are defined  at the center of the star, at $\xi=0$, while the remaining ones 
are specified  far from the star, at some point  $\xi=\xi_2$.} The boundary conditions for the metric components and the scalar field are 
\beq
\hM(0)=0,\quad  \nu(\xi_2)=0, \quad \hphi'(0)=0,\quad \hphi(\xi_2)=\hphi_\as. \label{boundaries}
\eeq
The boundary conditions for $\hM$ and $\hphi$ at the center just follow from the requirement of regularity  at $\xi=0$.  The condition on $\nu$ is in 
fact  arbitrary, since only $\nu'$ enters (\ref{eqs-r}) but not $\nu$ itself: our choice corresponds to defining the time coordinate as the proper time of 
a static observer at $\xi_2$. The ``asymptotic'' value $\hphi_\as$ corresponds to the minimum of the effective potential $\hV_{\rm eff}$ far from the star\footnote{Some situations require a more precise boundary condition, in which case  the exact value of $\phi$ at $\xi_2$ can be 
deduced from the asymptotic behaviour of the scalar field at large distances, which can be computed analytically.}.
Finally, the boundary condition for the pressure
 is 
\beq
\hP(0)=\hP_c\quad {\rm or}\quad \hP(\xi_2)=\hP_\infty \, , \nonumber
\eeq
depending on the problem: for a constant-density star it is convenient to choose $\hP(\xi_2)=-\hrho_\infty$ so that the solution is asymptotically de 
Sitter, while for the polytropic equation state,  the pressure  at the origin is fixed\footnote{In this case we also obtain asymptotically de Sitter 
solution.}.

%%%%%%%%%%%%%%%%%%%%%%%%%%%%%%%%%%%%%%%%%
\section{Chameleon}
\subsection{Description of the models}
In this section we consider chameleon scalar-tensor theories, introduced in \cite{Khoury:2003aq,Khoury:2003rn}. In these models, the effective 
mass of the scalar field becomes important inside matter so that the usual constraints on fifth-force interactions are evaded. To illustrate the 
behaviour of a chameleon field in a relativistic star, we will concentrate on a specific model, but our results should remain  qualitatively valid   for 
other models. 

We will consider the action (\ref{action}) with the 
potential
\be
V(\phi)=\frac{\mu^5}{\phi}, \nonumber
\ee
where $\mu$ is a mass parameter. 
We will also assume that $Q$ is positive. Moreover, as in the original proposal for the chameleon \cite{Khoury:2003aq,Khoury:2003rn}, we will use $Q\sim 1$ and $\mu\sim 10^{-3}$ eV  (this value corresponds to an upper bound for  $\mu$  in order  to satisfy the equivalence principle constraints \cite{Khoury:2003aq}).

If $Q\phi/M_P\ll 1$, which will be true for our configurations, 
the scalar field value corresponding to the minimum of the effective potential (\ref{Veff}) is given by
\be
\label{phimin-st}
\phi_{\rm min}=\sqrt\frac{M_P\mu^5}{Q(\trho-3\tP)}\, . \nonumber
\ee
Note that this is defined only if $\trho-3\tP>0$. Therefore, there is no local minimum at the center of a  constant energy density star with $
\Phi_*>5/18$.

For our numerical  study, we  have chosen $\e_0$ such that the asymptotic value of the scalar field is normalized to a value of order unity.
More precisely, we take \footnote{Any other choice is acceptable, but not so well adapted for numerics. For example, the choice $\epsilon_0=1$ leads to  values of the rescaled scalar field that are  very small fractions of unity, which is not very convenient for the numerical investigation.}
\be
\e_0=\sqrt{\frac{\mu^5}{QM_P\trho_\infty}} \nonumber
\ee
where $\trho_\infty$ is  the asymptotic energy density, so that the local minimum is given by
\beq
\hphi_{\rm min}=\sqrt{ \frac{\rho_\infty}{(1-3 w)\rho}}\, , \nonumber
\eeq
in rescaled units.
Asymptotically, one thus finds $\hphi_{\infty}=1/\sqrt{1-3w_\as}=1/2$ (for $w_\as=-1$).

As for the parameter $r_0$, it is convenient to choose it of the order of the star radius. In the case of the polytropic equation of state, we take $r_0= 
M_P \trho_\c^{-1/2}$, 
so that $\hrho_c=1$, as mentioned earlier.
 For the constant energy density stars, we will adopt the same prescription for a reference star characterized by $\Phi_*=0.165$. We will then 
compare this star to other stars, with the same radius but different central densities, by keeping $r_0$ fixed but by allowing the rescaled central 
density $\hrho_c$ to vary.

The rescaled potential can be written in the form
\be
\label{pot-r}
\hV = \frac{v_0}{\hphi},\qquad v_0=\frac{Q}{\e_0}\hrho_\as \, .
\ee
In rescaled units, taking $\e_1=1$,  the effective potential is thus 
\beq
\label{eff_pot-r}
\hV_{\rm eff}=\frac{v_0}{\hphi}+ \frac{1}{4\e_0^2}e^{4Q\e_0\hat\phi}(\hrho-3\hP).
\eeq
while the corresponding effective squared mass is
\beq
\label{meff2-r}
\hm_{\rm eff}^2=\frac{2 v_0}{\hphi_{\rm min}^3}+4Q^2  e^{4Q\e_0\hat\phi_{\rm min}}(\hrho-3\hP).
\eeq

Let us estimate the order of magnitude of the above parameters. A typical value for the density inside neutron stars is $\rho_\c\sim 10^{15} {\rm g}/{\rm 
cm}^3$, whereas the asymptotic density  is $\rho_\infty\sim 10^{-24} {\rm g}/{\rm cm}^3$, corresponding to the galactic density. 
With $Q\sim 1$ and $\mu\sim 10^{-3}$ eV, this leads to tiny values for our dimensionless  parameters: $
\epsilon_0\sim 10^{-19}$ and $v_0 \sim 10^{-21}$. These realistic parameters are too small  for the numerical precision that can be reached in 
practice and we have used parameters that are much bigger. 

%figure
\begin{figure}[ht]
\centering
\includegraphics[width=0.5\textwidth, clip=true]{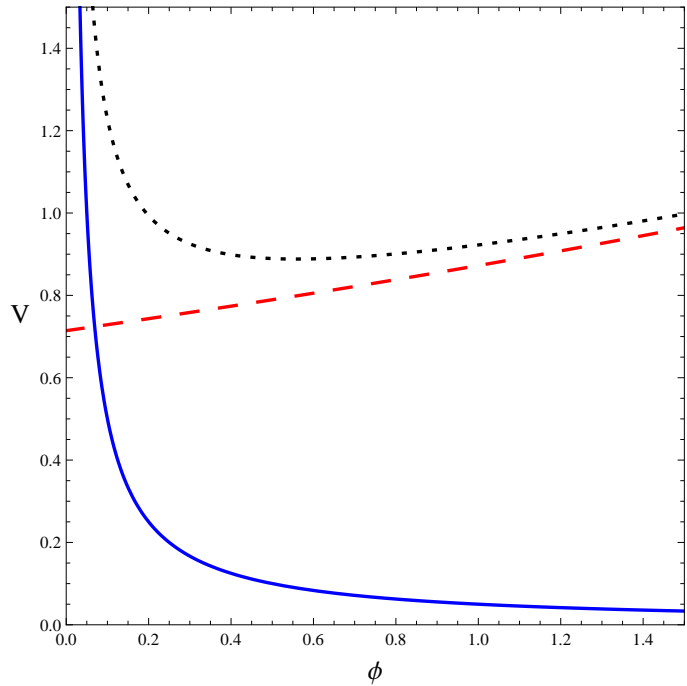}
\caption{The rescaled bare potentials (\ref{pot-r}), shown by solid blue, the matter part of  
the effective potential, shown by dashed red, and the effective potential for  
the chameleon model (\ref{eff_pot-r}), shown by dotted black line. The parameters are  
chosen as follows, $Q=1$, $v_0=0.05$, $\epsilon_0=0.05$, $\epsilon_1=1$ and $\hrho-3\hP=0.14$.
}
%\label{V}
\end{figure}
Asymptotically, the expression yields (\ref{meff2-r}) the tiny value
\be
\label{meff2-st}
m^2_{{\rm eff},\as} \simeq 2 v_0 (1-3w_\as)^{3/2}, \nonumber
\ee
where the second term is negligible for $\e_0\ll 1$. By contrast,  at the center of the star, the minimum is extremely small,
\beq
\hphi_{{\rm min}, c}= \sqrt{\frac{v_0\epsilon_0}{Q(1-3w_c)}}, \label{phiminc}
\eeq
while the effective mass becomes huge
\beq
m^2_{{\rm eff},c}  \simeq 2 \sqrt{\frac{Q^3(1-3w_c)^3}{v_0 \epsilon_0^3}}. \nonumber
\eeq
One can thus expect that the scalar field configuration will follow its local minimum (when it is defined) in the denser regions of the star.

\subsection{Numerical results}

Using the full system of equations combining Einstein's equations and the Klein-Gordon equation, we have computed numerically the profile  of 
the chameleon field, of the matter and of the geometry for two types of star configurations. 

Before discussing these two cases, let us comment about a subtlety concerning the asympotic behaviour of the geometry far from the star. In a 
realistic context,  the star configuration should be matched asymptotically  to a cosmological geometry, which is time dependent. Since we restrict 
our analysis to purely static configurations, we introduce instead some artificial matter which behaves like a cosmological constant far from the star. At 
large radius, the geometry thus approaches a Schwarzschild-de Sitter metric, which is static. We have checked that the details of the asymptotic 
matter do not modify the star configuration. 
Another possibility would be to introduce some non-trivial minimum in the chameleon potential.

\subsubsection{Constant energy density stars}
As a first (toy) example we considered a  constant energy density star. 
In order to avoid unnecessary numerical difficulties connected with a sharp transition of the density from the star to the surrounding medium, we 
introduced the following smoothed profile for the density:
\be
\hrho(\xi) = \frac{\hrho_c}{2} \left[1-\tanh \left(\frac{\xi -1}{\sigma }\right)\right]+\hrho_\infty, \label{eoscd}
\ee
where $\hrho_c$ is the rescaled density at the center of the star, $\hrho_\infty$ is the density of the surrounding medium far from the star and 
$\sigma$ is a ``smoothing'' parameter (with values between $10^{-12}$ and $10^{-2}$). It is easy to see, that for small enough $\sigma$ the density 
is approximately equal to $\hrho_c$ for 
$\xi\lesssim 1$ and $\hrho\simeq \hrho_\infty$ for $\xi\gtrsim 1$, so that $\xi\simeq 1$ is the radius of the star.  
The boundary condition for the pressure is, $\hP(\xi_2)=-\hrho_\infty$, so that we ensure that the de Sitter asymptotic behaviour  is recovered for our numerical 
solution. In this case the pressure is found as a function of $\xi$, and $\hP$ inside the star depends on the boundary condition $\xi_2$. This 
seemingly paradoxical situation is due to the artificial choice (\ref{eoscd}), but, as we have checked, the effect is tiny because the ratio between the 
star energy density and the asymptotic one is huge.

\begin{figure}[ht]
\centering
\includegraphics[width=0.5\textwidth, clip=true]{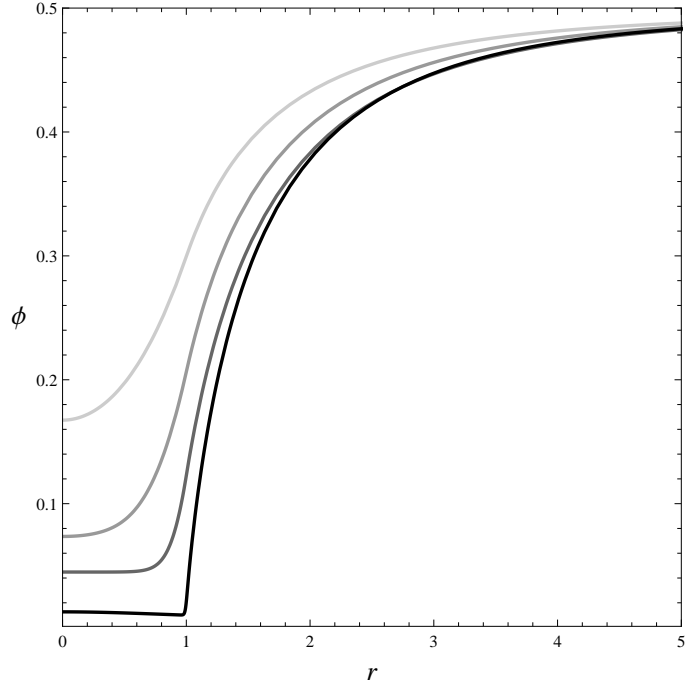}
\caption{
Profiles of the scalar field $\phi$ (in Planck units) as a function of radius of the star (in units of $r_0=M_P{\hat{\rho}_c  \tilde{\rho}_c}^{1/2}$) 
 for  constant density stars with the equation of state (\ref{eoscd}). The parameters are chosen as follows, $\sigma=0.01$, $Q=
\epsilon_1=1$, 
$ \epsilon_0=v_0=0.01$, and the rescaled densities of  the star are (from light gray to black), $\hat{\rho}_c=0.01$, $0.02$,  $0.05$, $1$.  
The values of the gravitational potential at the surface of the star are, respectively, $0.00168$, $0.00334$, $0.0083$ and $0.165$.
The increasing rescaled density corresponds to  an increasing physical density while the physical radius of the star  is
being fixed.}
\label{constant_chameleon}
\end{figure}

For a given coupling $Q$, the numerical system depends on  three independent parameters: $\hrho_c$, $v_0$ and $\epsilon_0$, while $\hphi_
\as=1/2$ and  the rescaled asymptotic density $\rho_\as$ is  determined by $\hrho_\as=v_0\e_0/Q$ according to (\ref{pot-r}). Since the 
backreaction of the scalar field on the star itself is very small, the compactness of the star depends only on the parameter $\hrho_c$. 
Consequently, for fixed parameters $v_0$ and $\e_0$, increasing the central density leads to stars that are more and more compact, and therefore 
more and more relativistic.  
 In Fig.~\ref{constant_chameleon}, we plot the profile of the scalar field for  stars with different central densities, corresponding to a gravitational 
potential ranging from  $0.00168$ to $0.165$.  As the central density increases, one sees how the scalar field profile evolves from a smooth 
configuration to a thin shell configuration, where  the scalar field abruptly jumps within a small range of radii. This is in agreement with the 
Newtonian result \cite{Khoury:2003rn} that 
$ \Delta r/\rs \approx (\phi_\as-\phi_c)/(6Q M_P\Phi_*)\approx \e_0/(12 Q \Phi_*)$. 
%figure

Essentially the thin-shell solution can be understood as follows. The solution tries to minimize the sum of the potential energy  
and of the gradient energy, with the constraints that $\phi'=0$ at $r=0$  and $\phi$ fixed asymptotically. As the energy density in the star  
increases, to be away from the minimum of the effective potential  becomes very costly energetically and, therefore, the field prefers to  
stay as much as possible close to its minimum.
At some point close to the star radius, however, the field starts to move away from its effective  minimum. In this region, the contribution from the potential on  
the right hand side of the Klein-Gordon equation becomes negligible with respect to the  matter dependent term, and the field ``evolves'' very 
quickly,  driven by this ``matter force" term. 
In the relativistic regime, analytical expressions describing the profile of the scalar field in the thin shell regime have been obtained  in 
\cite{Tsujikawa:2009yf}: they apply to  constant energy density stars and in the linearized approximation for the gravitational potential. 

\begin{figure}[ht]
\centering
\includegraphics[width=0.5\textwidth, clip=true]{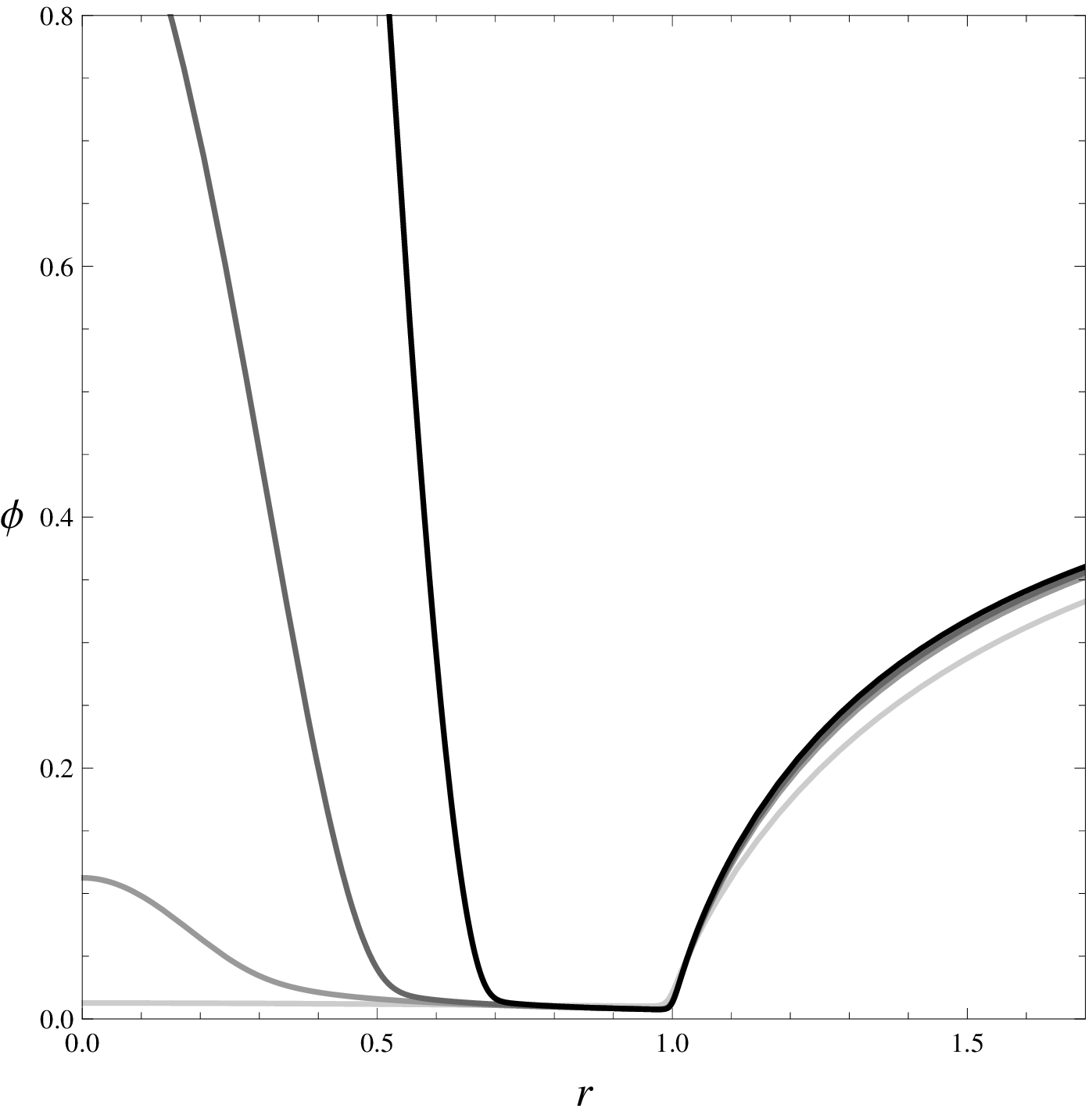}
\caption{
Profiles of the scalar field $\phi$ (in Planck units) as a function of radius of the star (in units of $r_0=M_P{\hat{\rho}_c \tilde{\rho}_c}^{1/2}$)  for 
stars with $\trho-3\tP<0$ in the central region. The rescaled densities of the star (from light gray to black): $\hat{\rho}_c=1$, $1.7$, $1.8$, $1.9$. The 
other parameters  are the same as in Fig.~\ref{constant_chameleon}. 
The values of the gravitational potential at the surface of the star are, respectively, $0.165$, $0.280$, $0.298$ and $0.318$.}
\label{negative_P}
\end{figure}

If one continues to increase the central energy density of the star, one ends up reaching the critical value for which $\hrho-3 \hP<0$ in the 
innermost central region of the star. The profile of the scalar field in the central region of the star is then radically altered, as one can see on Fig.~
\ref{negative_P}, because there is no local minimum for the scalar field in this region.

%figure

\subsubsection{Thin shell effect}
It is also instructive to see how much, depending on the density of the star,  the geometry outside the star is close to that expected in general 
relativity. 
 Numerically, one can estimate the post-Newtonian parameter $\tilde \gamma$ by comparing the coefficients of the terms in $1/r$ in the metric 
components $\tilde g_{tt}$ and $\tilde g_{rr}$. 

%figure
\begin{figure}[b]
\centering
\includegraphics[width=0.5\textwidth, clip=true]{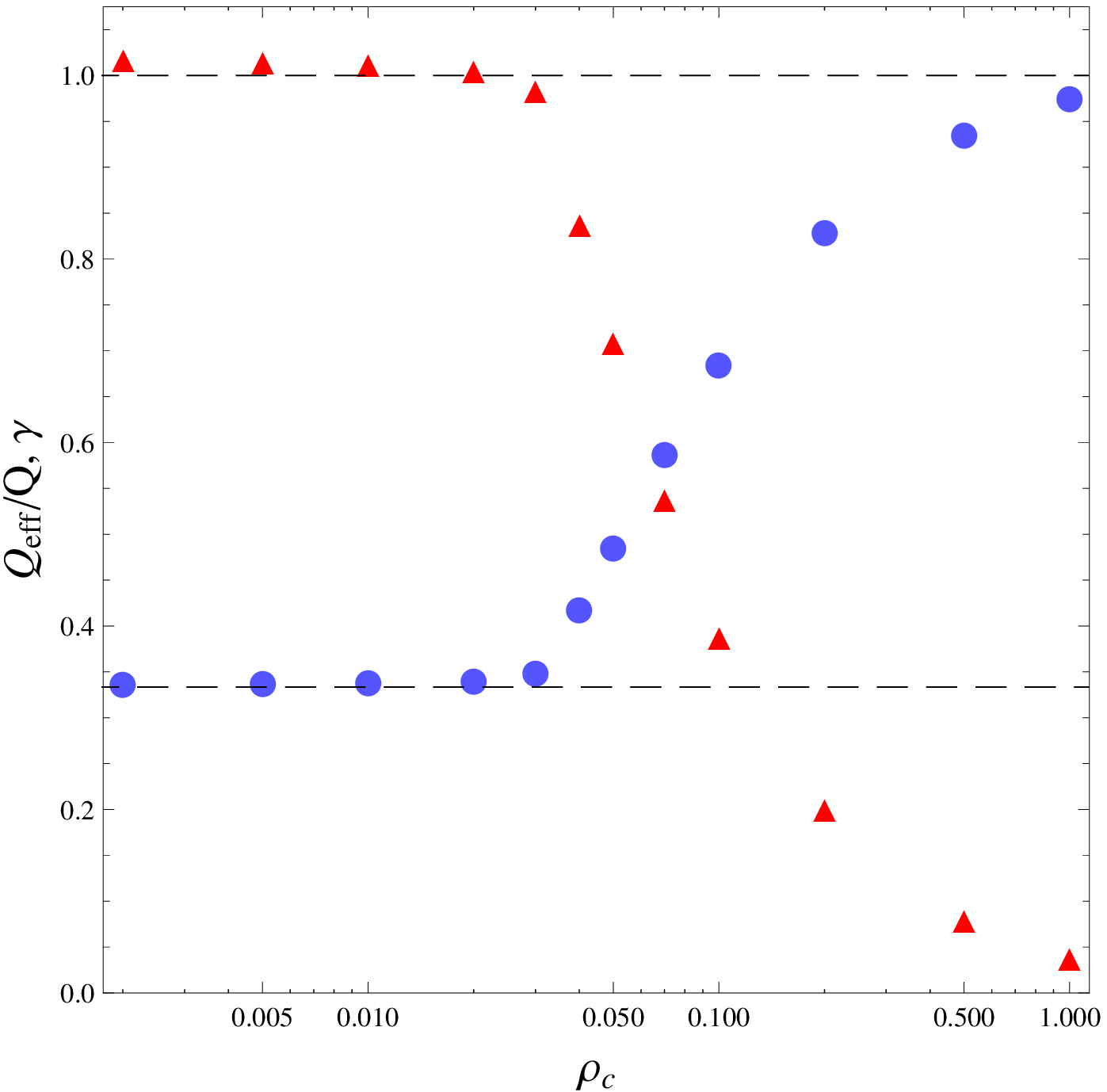}
\caption{Evolution of the numerical estimates for $Q_{\rm eff}/Q$  (red triangles) and for the post-newtonian parameter $\tilde\gamma$ (blue dots) 
as we increase the 
energy density of the star (constant energy density star). The ``bare'' coupling is here $Q=1/2$, which corresponds to 
$\tilde\gamma=1/3$ in the absence of screening. The other parameters are $\epsilon_0=10^{-2}$, $v_0=10^{-8}$ and $\xi_*=1$. 
}
\label{bending}
\end{figure}

Since we work in the Einstein frame, let us first discuss the metric components $g_{tt}$ and $g_{rr}$. One subtlety here is that the metric is 
asymptotically Schwarzschild-de Sitter rather than simply Schwarzschild and the asymptotic behaviours of the metric components are therefore
\beq
\label{metric_asymptotic}
e^\nu\approx e^{-\lambda} \approx 1-\frac{2GM}{r}-\frac{\Lambda}{3}r^2. 
\eeq
To estimate $M$ numerically, we have substracted from the metric components  the cosmological constant term with  
\be
\Lambda=\rho_\infty e^{4Q \phi_\as/M_P}+V_\as, \nonumber
\ee
where $\phi_\as$ is the asymptotic minimum of the scalar field and $V_\as\equiv V(\phi_\as)$. We have checked that the two metric components, in 
the Einstein frame, provide the same value $M$.
This means, in particular, that the backreaction due to the scalar field gradient is negligible outside the star.

In the Jordan frame, the metric components are $\tilde g_{\mu\nu}=\exp(2Q\phi/M_P)g_{\mu\nu}$. The asymptotic behaviour of the scalar field is of 
the form
\be
\phi\approx \phi_\infty+2\frac{GM}{r}Q_{\rm eff}e^{-m_{\rm eff} r}, \nonumber
\ee
where $Q_{\rm eff}$ is the effective coupling of the scalar field to the star, which is strongly suppressed in the thin shell regime. 
Combining this expression with (\ref{metric_asymptotic}),  it is easy to get
\be
\begin{aligned}
\tilde g_{tt} & \approx e^{2Q\phi_\as}\left[1-2\left(1-2Q  Q_{\rm eff}\right)\frac{GM}{r}-\frac{\Lambda}{3}r^2\right], \nonumber \\
\tilde g_{rr}^{-1} & \approx e^{-2Q\phi_\as}\left[1-2\left(1+2Q  Q_{\rm eff} \right)\frac{GM}{r}-\frac{\Lambda}{3}r^2\right]\, , \nonumber
\end{aligned}
\ee
where, for simplicity, we have neglected the exponential decay. 
The effective $\tilde\gamma$ parameter is therefore
\beq
\tilde\gamma=\frac{1-2 Q  Q_{\rm eff}}{1+2 Q  Q_{\rm eff}}\, . 
\eeq
The general relativistic value, $\tilde\gamma=1$, is thus recovered in the limit where the effective scalar charge $Q_{\rm eff}$ of the star is strongly 
suppressed, i.e. in  the thin shell limit. This is illustrated in Fig.~\ref{bending}, where we have estimated numerically the effective coupling $Q_{\rm 
eff}$ as well as the post-Newtonian parameter $\tilde\gamma$ for stars with increasing energy density.

\subsubsection{Polytropic relativistic stars}

Let us now consider relativistic stars with the more realistic equation of state (\ref{eos-R}). Although this particular equation of state 
is appropriate for the bulk of the star, it must be slightly modified  at the edge of the star for the following reasons. 
First of all, the equation of state (\ref{eos-R}) does not describe the exterior of the star.
However, in the case of the chameleon field it is crucial to have a non-zero background density,
therefore for small densities Eq.~(\ref{eos-R}) must be changed to include the background (galactic or extragalactic) matter in the model.
Moreover Eq.~(\ref{eos-R}) leads to a problematic behavior in the limit
$\rho\to 0$ (i.e. at the edge of the star), where the equation of state
becomes $\hP \propto \hrho^2$.
In this case, the relativistic Euler equation become ill-defined.
One can also see this problem when describing the dynamics of a perfect
fluid with $\hP \propto \hrho^2$ in terms of a scalar field:
the emergent metric for perturbations becomes singular when $\hrho\to 0$,
as it is explained in detail in \cite{Babichev:2007dw}.
We choose the following modification of the equation of state (\ref{eos-R}), 
\beq
\hrho=\hP+\left(\frac{\hP+\hrho_\infty}{0.39}\right)^{1/2},\label{eos-MOD}
\eeq
where we have introduced the (rescaled) asymptotic energy density $\hrho_\infty$, which is  an extremely small parameter with respect to the 
pressure and energy density in the star. 
It should be stressed for $\hP\gtrsim \hrho_\infty$ the modified equation
of state (\ref{eos-MOD}) takes the original form of the
equation of state (\ref{eos-R}), while for small $\hP$ the modified
equation of state allows for the cosmological term behavior
\footnote{In fact, our numerical integration shows that the system settles down at the cosmological term, $\hP=-\hrho$, at large $\xi$.}. 
Thus the modification of the equation of state only affects a tiny region
at the edge of the star. We also tried other
modifications for the equation of state, and we obtained the same results,
up to small differences around the edge of the star, as expected.
It is easy to see that for small $\hrho_\infty$ the ``fluid'' (\ref{eos-MOD}) is in the regime of a cosmological constant when $-\hP=\hrho\simeq\hrho_
\infty$.

Having specified the equation of state for the star and the surrounding matter (\ref{eos-MOD}), we also need to fix the boundary conditions. 
Apart from the four usual conditions (\ref{boundaries}), the boundary condition on the pressure is specified at the center of the star: $\hP(0)\simeq 
0.23$ (which corresponds to a realistic value, in the rescaled units). 
It is interesting to note, that the system ``chooses'' by itself the cosmological term-like behavior asymptotically, $\hP=-\hrho$, we do require any 
conditions on the density nor the pressure at infinity. This situation is in contrast to the constant density star configuration, where we imposed 
the ``correct'' asymptotic behavior at infinity specifying the boundary condition for $\hP$. 

\begin{figure}[ht]
\centering
\includegraphics[width=0.5\textwidth, clip=true]{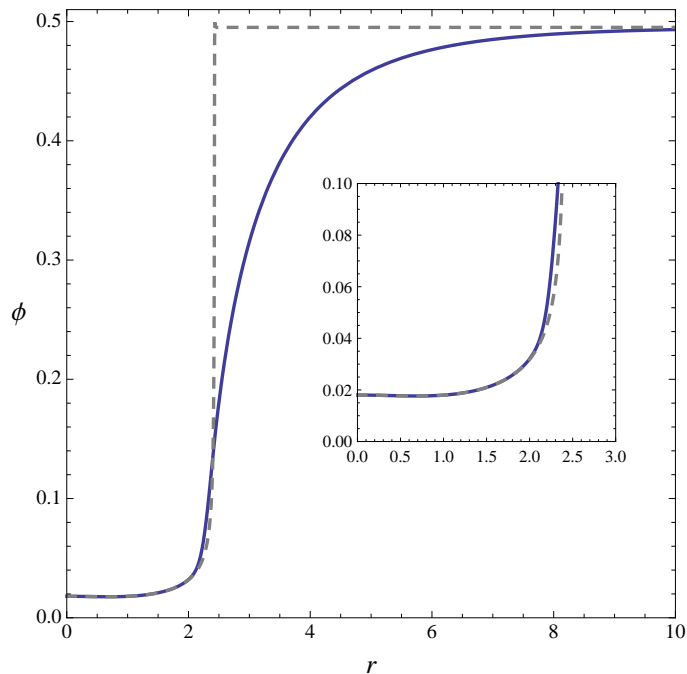}
\caption{Profile for the scalar field $\phi$ (in Planck units), shown by solid  
(blue) line, as a function of the radius (in units of $r_0=M_P\tilde{\rho}_c^{-1/2}$), for the equation of state (\ref{eos-MOD}) with $\epsilon_0=0.01$,  
$v_0=0.01$ and $\hrho_\as=10^{-4}$.  The value of the gravitational  potential, $\hM(\xi)/\xi$, at the surface of the star is $0.211$. The value $\phi_{\rm 
min} $ for the minimum of the effective potential is plotted by dashed  
(gray) line.}
\label{chameleon_realistic}
\end{figure}

The numerical solution for the scalar field profile is shown in Fig.~ \ref{chameleon_realistic} for the parameters $Q=
\epsilon_1=1$, $\epsilon_0=0.01$, $v_0=0.01$, $\rho_0=10^{-4}$. The minimum of the effective potential is indicated by a dashed line. In the 
central part of the star, the scalar field strictly follows this minimum. Note that, as the radius increases, the minimum first increases and then 
decreases. This non-monotonous evolution is the consequence of the non-monotonous evolution of the quantity $\trho-3\tP$, which was  pointed 
previously. Slightly before reaching the radius of the star (delimited by the quasi-vertical dashed line), the scalar field starts to move away from its 
minimum, with a steep gradient, and asymptotically evolves toward its asymptotic minimum far from the star.
We have also studied  numerically configurations with other sets of
parameters, namely in the range $\epsilon_0\sim 0.1-10^{-3}$, $v_0\sim
0.1-10^{-3}$, $\rho_0\sim 10^{-2}-10^{-4}$, and found a similar
behaviour, while the value of the scalar field at the center of the
star agrees with the estimate (\ref{phiminc}).

As mentioned earlier,  the profiles for the energy density and pressure of the star are almost unchanged with respect to those for the corresponding star in pure general relativity (i.e. without scalar field), presented in Fig.~\ref{star}. This can be understood by noting that  the contribution  of the scalar field in the first three equations (\ref{eqs-r}) can be neglected, giving  then the same  equations as in general relativity.
The change in the total  neutron star mass is less than one percent.

%figure

\section{$f(R)$ gravity}
We now turn to  $f(R)$ gravity, restricting ourselves to the conventional metric formulation. Some of our results on relativistic stars have already 
appeared
 in a separate publication \cite{BL1}.  Here, we  present more details and stress the strong analogy of these models with the chameleon models. 

\subsection{The models}
The $f(R)$ models are usually described  by  an action of the form
\beq
\label{jordan}
S=\frac{M_P^2}{2}\int d^4x \sqrt{-\tg}\,f(\tR) + S_m[\Phi_m;\tg_{\mu\nu}], 
\eeq
where the matter is minimally coupled to the  metric $\tg_{\mu\nu}$, with corresponding Ricci tensor 
$\tR_{\mu\nu}$ and scalar curvature $\tR$. It is convenient to reexpress $f$ in the form
\beq
f(\tR)=R_0 \F(x), \qquad  x\equiv\frac{\tR}{R_0} \nonumber
\eeq
 where $\F$ is a dimensionless function and $R_0$ is some parameter.

In contrast with many papers on $f(R)$ theories, in particular \cite{Frolov} and \cite{KM1}, we choose here to reexpress the model as a scalar-
tensor theory in the so-called Einstein frame. 
The two formulations are of course equivalent (at least at the classical level), but the Einstein frame is useful to compare directly the behaviour of 
$f(R)$ theories with that of  chameleon models, discussed in the previous section. 

By introducing the scalar field 
\beq
\label{phi_R}
\phi=\sqrt{\frac{3}{2}} M_P \, \ln f_{,\tR}=\sqrt{\frac{3}{2}} M_P \, \ln \F'(x)
\eeq
 and the metric 
 \beq 
 g_{\mu\nu}=\Omega^{-2}\, \tg_{\mu\nu}, \quad \Omega^{-2}=f_{,\tR}=\exp[\sqrt{\frac23}\, \phi/M_P] \nonumber
 \eeq
the action (\ref{jordan}) can be reexpressed in the form (\ref{action}) with 
the coupling 
\beq
Q=-1/\sqrt{6}, \nonumber
\eeq
and the potential
\beq
\label{potential}
V=M_P^2\, \frac{\tR f_{,\tR}-f}{2f_{,\tR}^2},
\eeq
which can be expressed in terms of $\phi$ by inverting the definition (\ref{phi_R}) of  $\phi$ as a function of  $\tR$. 

Note that, in contrast to the chameleon case discussed in the previous section, the coupling $Q$ is here negative. We could have chosen a 
positive value by changing the sign of $\phi$, but we have kept the usual convention. When comparing the results of this section with the previous 
one, it is useful to keep in mind that the signs of $Q$ and $\phi$ can be simultaneously changed.

%figure
\begin{figure}[ht]
\centering
\includegraphics[width=0.5\textwidth, clip=true]{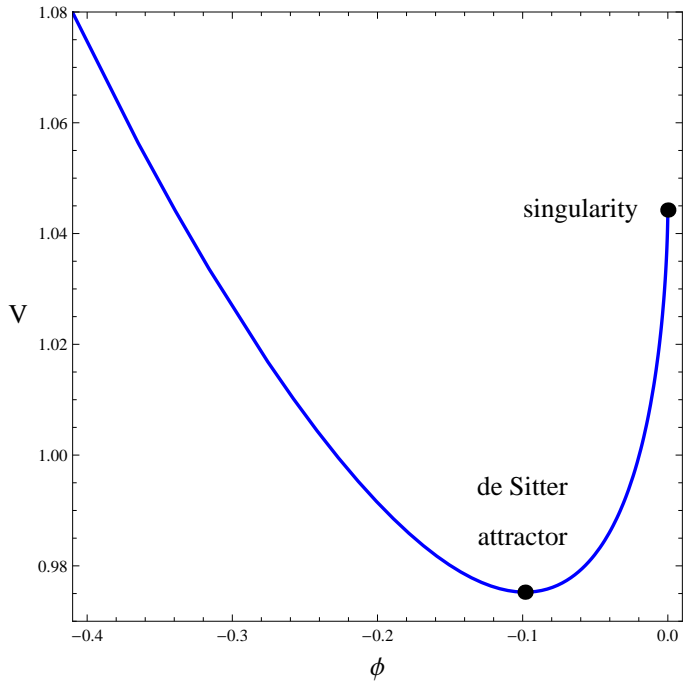}
\caption{Potential $V$ (in units of $M_P^2 \R0$) as a function of $\phi$ (in Planck units) for $n=1$ and $\x1=3.6$.
The lower black dot corresponds to the de-Sitter attractor while the upper-right dot shows the curvature singularity.}
\label{V}
\end{figure}

The function $f(R)$ is severely constrained by observations. For the present discussion, we focus our attention on the specific class of models 
introduced  by Starobinsky \cite{Starobinsky:2007hu},
\beq
\label{staro}
f(\tR)=\R0\left[x-\lambda 
\left(1-\left( 1+x^2\right)^{-n} \right)\right], \quad x\equiv \frac{\tR}{\R0}\, .
\eeq
The asymptotic de Sitter solution $\tR_\infty\equiv \x1 \R0$ corresponds to a minimum of $V$ and depends on the parameter $\lambda$. For 
practical purposes, it is simpler to choose $\x1$ as the parameter and then express $\lambda$ as a function $\x1$, with the restriction that $\x1$ 
must be chosen such that it corresponds to a stable minimum of the potential. Details and explicit expressions can be found in the appendix. The 
potential for the scalar field in the model with $n=1$ and $\x1=3.6$ is plotted in Fig.~\ref{V}.

\subsection{High curvature regime}
Inside the star,  the curvature is much higher than the cosmological curvature $\R0$, i.e.
$x\gg 1$. It is instructive to consider the asymptotic expansion of the potential and its derivatives in order to understand intuitively the behaviour of 
the scalar field, noting that the infinite curvature limit corresponds to $\phi\rightarrow 0^-$.

Using the asymptotic expansion 
\beq
\F(x)\simeq x-\lambda +\lambda x^{-2n}, \qquad (x\gg 1),
\label{f_approx}
\eeq
one finds 
\beq
{\cal V}(x)\equiv \frac{V}{M_P^2 R_0}\simeq \frac{\lambda}{2}\left[1-(1+2n)x^{-2n}\right],     \nonumber
\eeq
which shows that the amplitude of the potential remains {\it finite} when $x\gg 1$, i.e. when $\phi$ approaches $0$. This is quite different from the 
chameleon model discussed in the previous section and this can cause some worry that the singularity can be easily accessible for high enough 
curvatures, as discussed in \cite{Frolov}. 

However, the derivative of the potential still goes to infinity in this limit and this property prevents the scalar field to reach the singularity. 
Indeed, the derivative of the effective potential is given by (see appendix)
\beq
\frac{d V_{\rm eff}}{d\phi}\simeq \sqrt{\frac23}\mP \R0 \frac{x}{2}-\frac{1}{\sqrt{6}\, \mP}(\trho-3\tP), \nonumber
\eeq
which implies the existence of a minimum 
\beq
\label{minimum}
x_{\rm min}\simeq \frac{\trho-3\tP}{\mP^2 \R0}, \qquad \phi_{\rm min}=-\sqrt{6}\lambda n x_{\rm min}^{-2n-1}.
\eeq
 at least if the matter term $\trho-3\tP$ is positive. It can be checked that this is indeed a minimum by computing the second derivative of the 
effective potential. This leads to a positive effective square mass (see appendix),
 \beq
m_{\rm eff}^2\simeq \R0\frac{x_{\rm min}^{2n+2}}{6\lambda n (2n+1)}, \nonumber
\eeq
which becomes very large at high curvature.

 As  the energy  
density in the star increases toward the center, one expects that the scalar field  will  
be closer and closer to the singularity (without reaching it however),  
simply because the minimum of the effective potential is closer and  
closer to the singularity. This is confirmed by our numerical calculations, which we dicuss below.

\subsection{Numerical results}
We have integrated numerically the system of equations (\ref{eqs-r}),
together with (\ref{eos}), for Starobinsky's model  (\ref{staro}) with $n=1$.  Our parameters are $\epsilon_0=\epsilon_1=1$ and the rescaled 
potential is 
\beq
\hV=v_0 \, \V \, ,\qquad  v_0=r_0^2 \R0=\frac{M_P^2\R0}{\trho_c},     \nonumber
\eeq
where the explicit expression for $\V$ in the case $n=1$ is given by (\ref{Vn1}) in the appendix.

Here, the parameter $v_0$ corresponds to  the ratio between the energy density at infinity (i.e. the cosmological energy density) and the energy 
density at the center of the star. 
Realistic values of this parameter are thus extremely small, typically
$v_0\sim 10^{-40}$,  and are numerically challenging to reach, because
the scalar field value at the center is proportional to $v_0^3$, as can be
seen by comparing  the definition of $v_0$ with the minimum defined in
(\ref{minimum}) in the case $n=1$. 
We have performed our numerical calculations in the range $v_0\sim
10^{-1} - 5\times 10^{-5}$ and checked that the solution at the center
behaves as expected from our analytical analysis, Eq.~(\ref{minimum}). 
We are thus confident that this should hold for smaller values of $v_0$.
As in the case of the chameleon model, the profiles for the density and pressure only slightly differ from the GR results.

\begin{figure}[t]
\centering
\includegraphics[width=0.5\textwidth, clip=true]{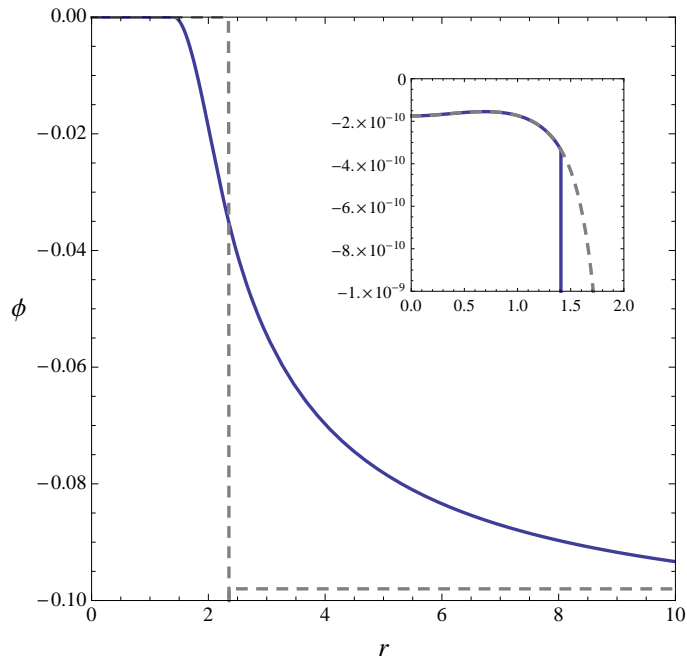}
\caption{Profile of  the scalar  field $\phi$ (in Planck units), shown by solid (blue) line, 
as a function of the radius (in units of $M_P\tilde{\rho}_c^{-1/2}$), 
for the model (\ref{staro}) with $n=1$, $\x1=3.6$ (see appendix for the definition of $\x1$) and $v_0=10^{-4}$. 
The value $\phi_{\rm min}$ for the minimum of the effective potential is plotted by dashed (gray) line.  }
\label{field}
\end{figure}

The profile of the scalar field inside and outside the star is plotted in Fig.~\ref{field}.
As is clear from the figure, the scalar field  tends to interpolate between an extremely high density regime, inside the star, and a very low density 
regime, outside the star. 
 This behaviour is quite analogous to that of the chameleon model discussed in the previous section and the solution that we obtain is analogous 
to  the thin-shell solution. 
What differs from the usual chameleon models is the presence of the singularity. In the very high density regime, the scalar field is very close to the 
singularity because the minimum of the effective potential is itself very close to the singularity. To give an analogy, it is like the scalar field is 
following a track very near a precipice, but because the effective mass is very high in this regime, the scalar field is securely attached to the track 
and does not fall into the nearby precipice.

These numerical results, previously presented in~\cite{BL1},  contradict the results of \cite{KM1}, where it was claimed that stars with a gravitational 
potential larger than a critical value $\Phi_{\rm max}\approx 0.1$ could not be constructed. The reason advocated in \cite{KM1} was the presence 
of the singularity, as stressed previously in \cite{Frolov}. 
However, as we showed in~\cite{BL1}, the analytical arguments in \cite{Frolov} and \cite{KM1} supporting this interpretation were not valid
\footnote{Note also that, obviously,  the problem encountered in \cite{KM1}
cannot be explained by the sign of the trace of the energy-momentum tensor  since their gravitational potential is well below the critical value 
(\ref{Phi_critical}).}.  
 We also  indicated that the exploration of the scalar field profile was numerically challenging for small values of $v_0$, because the distance 
between the scalar field value and the singularity is proportional to $v_0^3$. 
Since the value used in \cite{KM1} was $v_0\sim 10^{-6}$, one could strongly suspect  that the results obtained in \cite{KM1} were  probably due  
to a numerical instability.
This explanation was confirmed subsequently in \cite{Upadhye:2009kt}, which tried to reproduce the  numerical analysis as \cite{KM1}, using the 
same technique, namely the shooting method. Indeed, they were able to construct highly relativistic stars, but with a parameter $v_0\sim 10^{-2}$.

Although we show only our numerical  results in the case of the polytropic equation of state, we have also performed similar numerical integrations 
for constant energy density stars. We have obtained quite similar results as in the chameleon model. For instance, we have been able to  obtain 
numerical solutions up to the value  $\Phi_*=0.33$ for the gravitational potential \footnote{Note that a similar value for the maximum gravitational 
potential, $\Phi_*=0.345$, was obtained in \cite{Upadhye:2009kt}.} (with the parameters $\epsilon_1=\epsilon_0=1$, $v_0=0.01$).

\subsection{``Cured'' f(R) gravity}

\begin{figure}[b]
\centering
\includegraphics[width=0.5\textwidth, clip=true]{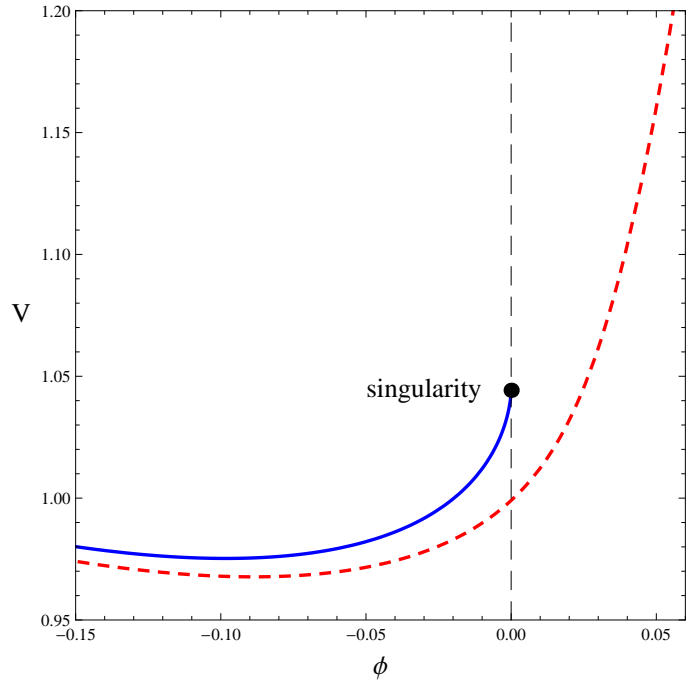}
\caption{Potential $V$ (in units of $M_P^2 \R0$) as a function of $\phi$ (in Planck units) for $n=1$ and $\x1=3.6$.
The blue curve corresponds to the original Starobinsky cosmological  
model with the curvature singularity at $\phi=0$. The red curve  
represents the ``cured'' Starobinsky model with $\sigma=10^{-3}$.}
\label{V_cured}
\end{figure}

Although the existence of neutron stars is, so far\footnote{It would be also worth considering a supernova explosion and the birth of a neutron star 
in the context of $f(R)$ theories. The present work deals only with static solutions and not how they could be reached in an evolutionary process.}, 
compatible with the $f(R)$ models we have just investigated, the situation is not the same in cosmology. It has indeed been noticed in  
\cite{Starobinsky:2007hu,Tsujikawa:2007xu,Appleby:2008tv} that the behaviour of these theories is pathological when one considers their past 
cosmological evolution, with the generic presence of a singularity in the not so distant past. 
As a way to ``cure''  this  pathology at high curvature, it has been suggested to add 
$R^2$  term\footnote{Another way to avoid a pathological behavior would be to construct $f(R)$ model separating the singularity with infinite 
energy barrier, as it was done in \cite{Miranda:2009rs}. It was argued, however, that this particular model does not satisfy local gravity constraints 
and SDSS data \cite{delaCruzDombriz:2009xk}. }. 
 This extra term radically changes the behaviour of the potential at very high curvature, as illustrated in 
Fig.~\ref{V_cured} for the model
\beq
\label{cured}
f(\tR)=\R0\left[x-\lambda 
\left(1-\left( 1+x^2\right)^{-n} \right)+\sigma\, x^2\right], \quad x\equiv \frac{\tR}{\R0}\, .
\eeq
One sees that the corresponding potential has no  singularity at $\phi=0$ and the scalar field can now take positive values, corresponding to high 
curvatures. 

The extra $R^2$ term has the additional advantage that it can drive inflation in the early Universe (see e.g.  \cite{Nojiri:2007as,Nojiri:2007cq}), as 
in the pioneering Starobinsky model \cite{Starobinsky:1980te}. 
Recently, however, it was argued in \cite{Appleby:2009uf} that one cannot relate smoothly the inflation epoch with the present acceleration epoch 
with the above model (\ref{cured}). 
As an example of a viable transition between  two accelerating solutions,   the model
\beq
\label{ABS}
f(\tR)=(1-c) \tR+c\, \epsilon\ln \left[\frac{\cosh (\tR/\epsilon-b)}{\cosh b}\right]+\frac{\tR^2}{6M^2}, \quad \epsilon\equiv \frac{R_0}{b+\ln(2\cosh b)}
\eeq
with $0<c<1/2$, 
was proposed in \cite{Appleby:2009uf}, where the new mass scale $M$ must be of the order $10^{12}$ GeV to reproduce the standard predictions 
of inflation.
\begin{figure}[ht]
\centering
\includegraphics[width=0.5\textwidth, clip=true]{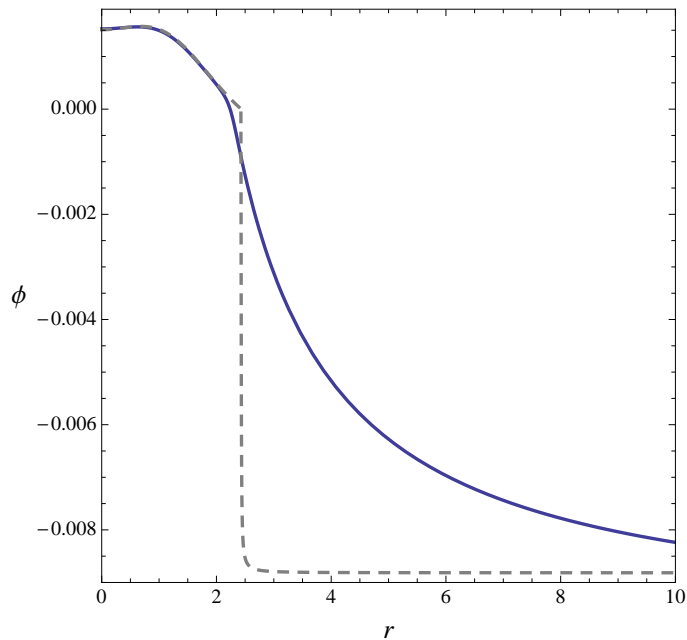}
\caption{Profile of the scalar field  in the  model (\ref{ABS})  with the parameters
$\sigma=2\times10^{-6}$, $c=0.4$, $b=4$, $v_0=10^{-3}$.}
\label{AB}
\end{figure}
In the high curvature regime ($R\gg R_0$), one finds
\beq
f(\tR)\approx \tR-cR_0+c\, \epsilon\,  e^{2b} \, e^{-2\tR/\epsilon}+\frac{\tR^2}{6M^2}\, .    \nonumber
\eeq
The interior of a  neutron star corresponds to the high curvature regime, but the curvature is many orders of magnitude smaller than $M^2$. In this 
intermediate regime $R_0\ll \tR\ll M^2$, the scalar field behaves
like
\beq
\label{phi_ABS}
\frac{\phi}{M_P}=\sqrt\frac32 \ln f_{,\tR}\approx \sqrt\frac32\left[-2 c\,  e^{2b} \, e^{-2\tR/\epsilon}+\frac{\tR}{3M^2}\right].
\eeq
The two terms on the right hand side are extremely small. However, the first term decays exponentially as the curvature increases and the second 
term will typically dominate at the center of a neutron star. 

We plot on Fig.~\ref{AB} the profile of the scalar field in this model  with small values for $v_0$ and $\sigma=R_0/(6M^2)$, although not small 
enough to be realistic. Once again, we observe that the scalar field follows the minimum of its effective potential in the core of the star. Near the 
boundary of the star, it deviates from the local minimum  and converges towards the asymptotic minimum outside the star. The scalar field is 
negative deep inside the star because the second term on the right hand side of (\ref{phi_ABS}) dominates, whereas it is positive near the 
boundary and outside the star. 

\section{Discussion and conclusion}
We have studied in this work  relativistic stars in scalar tensor and $f(R)$ theories that use the chameleon mechanism. The behaviour of the scalar 
field is extremely similar in the two types of models. The main properties are the following. 

Deep inside the star, the scalar field follows very closely the minimum of its effective potential (if it exists). As a consequence, if the minimum 
changes as a function of the radius, the scalar field will closely follow it.
 If $\trho-3\tP<0$, which can occur for instance in the central region of very compact stars with constant energy density, there is no minimum for the 
effective potential. It is however possible to find numerical solutions for constant energy stars with both $w<1/3$  and $w>1/3$ regions, but only up 
to some maximum gravitational potential. We have  given qualitative arguments to show that one expects tachyonic instabilities if most of the star is 
characterized by $w>1/3$.  

As already emphasized in our previous work \cite{BL1}, our results invalidate the claim that highly relativistic stars cannot be constructed in $f(R)$ 
theories. Numerically, the construction of a relativistic stars in $f(R)$ gravity is a  challenging task because  the scalar field value is extremely close 
to the singularity in the center of the star.   We have also constructed relativistic stars in the so-called ``cured'' $f(R)$ models, which have been 
advocated to solve a cosmological singularity problem. 
  
We have also illustrated how the screening effect of the chameleon mechanism manifests itself for relativistic  stars, by computing numerically the 
effective coupling of the star as well as the post-Newtonian parameter $\tilde\gamma$.

\vskip 1cm
{\bf Note:}
 while this paper was being completed, another paper~\cite{Cooney:2009rr} on relativistic stars in $f(R)$ theories appeared on the arXiv, although 
in a different context.

\vskip 1cm

\begin{acknowledgments}
We would like to thank Nathalie Deruelle, Gilles Esposito-Farese, Andrei Frolov, Eric Gourgoulhon, Kei-ichi Maeda, J\'er\^ome Novak, Ignacy 
Sawicki, Alexei Starobinsky, Shinji Tsujikawa and Riad Ziour for very instructive discussions.
We also thank Tomohiro Harada and Shinya Okubo for pointing out an incorrect statement in a previous version of this manuscript.
The work of E.B. was supported by the EU FP6 Marie Curie Research and Training Network UniverseNet (MRTN-CT-2006-035863).
 \end{acknowledgments}

\appendix
\section{}

\subsection{General formulas for $f(R)$ models}
For $f(\tR)$ models, the  potential $V$, introduced in (\ref{potential}), can be written as
\beq
\label{V_app}
V=M_P^2\, \frac{\tR f_{,\tR}-f}{2f_{,\tR}^2}=M_P^2 R_0 \V(x), \qquad \V(x)\equiv \frac{x \F'(x)-\F(x)}{2\F'(x)^2},
\eeq
where $f(\tR)=R_0 \F(x)$ and $x\equiv \tR/R_0$. 
The associated scalar field is defined by
\beq
\label{phi_app}
\phi=\sqrt{\frac{3}{2}} M_P \, \ln f_{,\tR}=\sqrt{\frac{3}{2}} M_P \, \ln\F'(x)
\eeq
Combining (\ref{V_app}) and (\ref{phi_app}), one finds for the first and second derivatives of the potential $V$ in terms of 
$\phi$ the following expressions:
\bea
\label{dV}
& &\frac{dV}{d\phi}=
\sqrt{\frac{2}{3}}M_P\,  \frac{2f-\tR f_{,\tR}}{2f_{,\tR}^2}\,, \\
& & \frac{d^2V}{d\phi^2}=\frac{1}{3f_{,\tR\tR}} 
\left[ 1+ \frac{\tR f_{,\tR\tR}}{f_{,\tR}}-\frac{4ff_{,\tR\tR}}{f_{,\tR}^2} 
\right]\,.
\label{d2V}
\eea

\subsection{Starobinsky's models}
Let us now restrict ourselves to  the specific class of models introduced  by Starobinsky\cite{Starobinsky:2007hu},
\beq
f(\tR)=\R0\left[x-\lambda 
\left(1-\left( 1+x^2\right)^{-n} \right)\right], \quad x\equiv \frac{\tR}{\R0}\, .  \nonumber
\eeq
Substituting this expression into (\ref{dV})  and solving $V'=0$ yields the minimum of $V$  corresponding to the asymptotic de Sitter solution $\tR_
\infty\equiv \x1 \R0$. It is convenient to express the parameter $\lambda$ in terms of $\x1$:
\beq
\lambda=\frac{\x1(1+\x1^2)^{n+1}}{2\left[ (1+\x1^2)^{n+1}-1-(n+1)\x1^2 \right]}\,.  \nonumber
\eeq
In terms of $\x1$,  the minimum of the potential  is given by
\beq
V_\infty=V(\phi_\infty)=\frac14 M_P^2 \R0 \x1\frac{(1+\x1^2)^{n+1}-1-(n+1)\x1^2}{(1+\x1^2)^{n+1}-1-(2n+1)\x1^2} \nonumber
\eeq
for
\beq
\phi_\infty=\sqrt{\frac{3}{2}} M_P \, \ln\left[1-\frac{n \x1^2}{(1+\x1^2)^{n+1}-1-(n+1)\x1^2}\right].\nonumber
\eeq
At the minimum, the effective squared mass is
\beq
m_\infty^2=\frac{d^2V}{d\phi^2}(\phi_\infty)= \frac{4 V_\infty}{M_P^2}
\frac{\left[\left(\x1^2+1\right)^{n+2}-\left(2
   n^2+3 n+1\right) \x1^4-(n+2) \x1^2-1\right]}{3 n \x1^2 \left[(2 n+1)
   \x1^2-1\right]}   \,. \nonumber
\eeq
The parameter $\x1$ must be chosen so that $m_\infty^2>0$.

Let us now consider the regime where the curvature is much higher than the cosmological curvature $R_0$, i.e.
$x\gg 1$. In this regime,
\bse
\label{f_approxA}
\begin{align}
f(\tR)&\simeq R_0\left(x-\lambda +\lambda x^{-2n}\right),\\
 f'(\tR)&\simeq 1-2\lambda n x^{-2n -1}, \\
f''(\tR)&\simeq \frac{2}{R_0}\lambda n(2n+1)x^{-2n-2},
\end{align}
\ese
and the scalar field behaves like
\beq
\label{phi_high}
\sqrt{\frac23}\frac{\phi}{\mP}\simeq -2\lambda n x^{-2n-1} \qquad (x\gg 1).\nonumber
\eeq
Substituting   in (\ref{dV}) the above expressions, one finds that the derivative of the effective potential, whose general expression can be found  in 
(\ref{dVeff}),
is given by  
\beq
\frac{d V_{\rm eff}}{d\phi}\simeq \sqrt{\frac23}\mP R_0 \frac{x}{2}-\frac{1}{\sqrt{6} \mP}(\trho-3\tP)\qquad (x\gg 1)\, . \nonumber
\eeq
The minimum of the effective potential is thus determined by
\beq
x_{\rm min}\simeq \frac{\trho-3\tP}{M_P^2 R_0}, \nonumber
\eeq
and exists only if the matter term $\trho-3\tP$ is positive. 
Substituting in (\ref{d2V}) the expressions (\ref{f_approxA}), one also finds that the effective mass at the minimum  behaves like
\beq
m_{\rm eff}^2\approx \frac{R_0}{6\lambda n (2n+1)} x_{\rm min}^{2n+2}.\nonumber
\eeq

In the particular case $n=1$, one can  invert the relation between $\phi$ and $\tR$ and obtain an  explicit formula expressing $x$ as a function of $
\phi$, which we do not write down because it is rather ugly. 
Substituting  this relation 
\beq
\label{Vn1}
\V(x)=\frac{\lambda \left(x^2-1\right) \left(x^3+x\right)^2 }{2 \left(x^4+2 x^2-2 \lambda  x+1\right)^2}
\eeq
thus gives an explicit expression for the potential as a function of $\phi$.

\end{document}